\documentclass[12pt]{article}

\usepackage{latexsym,amsmath,amssymb,theorem,epsfig}
\usepackage{hyperref}

\DeclareFontFamily{OT1}{rsfs10}{}
\DeclareFontShape{OT1}{rsfs10}{m}{n}{ <-> rsfs10 }{}
\DeclareMathAlphabet{\mathscript}{OT1}{rsfs10}{m}{n}

\numberwithin{equation}{section}

\newcommand{\tr}{\text{tr}}
\newcommand{\pt}{\partial}

\def\a{\alpha}
\def\b{\beta}
\def\g{\gamma}

\def\d{\delta}

\def\z{\psi}
\def\k{\kappa}
\def\l{\lambda}
\def\m{\mu}

\def\q{\theta}
\def\r{\rho}
\def\s{\sigma}

\def\z{\zeta}

\def\F{\Phi}
\def\G{\Gamma}

\def\L{\Lambda}

\def\cO{{\mathcal O}}

\def\gsim{ \lower .75ex \hbox{$\sim$} \llap{\raise .27ex \hbox{$>$}} }
\def\lsim{ \lower .75ex \hbox{$\sim$} \llap{\raise .27ex \hbox{$<$}} }
\def\be{\begin{equation}}
\def\ee{\end{equation}}
\def\bea{\begin{eqnarray}}
\def\eea{\end{eqnarray}}

\def \td {\tilde}

\def \ha {{1 \ov 2}}

\def \sql {{\sqrt{\l}}\ }
\def \del{\partial}
\def \a {\alpha}
\def \aa {{\a'}}
\def\ov{\over}
\def \ci {\cite}

\def \foot {\footnote}
\def \bi{\bibitem}
\def\la{\label}\def\foot{\footnote}\newcommand{\rf}[1]{(\ref{#1})}

\def \adss {$AdS_5 \times S^5\ $}

\def \a {\alpha }

\def \N  {{\cal N}}
\def \adt  {{$AdS_2\times S^2$\ }}
\def \cO {{\cal O}}

\theoremstyle{plain}

{\theorembodyfont{\rmfamily} }

\def \ed {\end{document}}

\begin{document}

\begin{titlepage}

\vspace{-1cm}


\vspace{-5cm}

\title{
  \hfill{\small Imperial-TP-AT-2014-03  }  \\[2 em]
   {\LARGE   
    The $1/N$  correction in  the D3-brane description of \\ circular Wilson loop
   at strong coupling
}
\\[1em] }
\author{
   E. I. Buchbinder
     \\[0.1em]
   {\it \small School of Physics M013, The University of Western Australia,}\\
       {\it \small 35 Stirling Highway, Crawley W.A. 6009, Australia}\\[0.5em]
A. A. Tseytlin\footnote{Also at Lebedev Institute, Moscow}
\\[0.1em]
   {\it \small The Blackett Laboratory, Imperial College, London SW7 2AZ, U.K.}
   }
   
\date{}
\maketitle
\begin{abstract}
We compute  the one-loop  correction to   the probe D3-brane  action  in  $AdS_5 \times S^5$ 
 expanded around   the classical Drukker--Fiol solution   ending on a circle  at the boundary. 
 It is  essentially   the  logarithm  of  the one-loop   partition function  of an Abelian $\N=4$ 
  vector multiplet in \adt geometry. 
 This one-loop  correction   should be 
  describing  the subleading  $1/N$  term in  the expectation value  of 
circular Wilson loop   in  the totally symmetric  rank $k$  representation  in  $SU(N)$   SYM theory 
 at strong coupling. 
 In the limit    $ 1\ll k\ll N $   when 
  the  circular Wilson loop  expectation values 
 for the symmetric representation   and   for  the product of  $k$  fundamental 
representations   are expected to  match  we find that 
 this one-loop     D3-brane correction     agrees  with  the  gauge  theory result
 for the $k$-fundamental  case. 
\end{abstract}

\thispagestyle{empty}

\end{titlepage}
\def \iffa {\iffalse}
\def \tr  {{\rm tr\,}}
\def \F {{\cal F}} 
\def \smm {{\rm sym} }  \def \edd {\end{document}}

\def \N {{\cal N}}
\def \te {\textstyle}
\def \ha  {{\te {1\ov 2}} }

\def \be {\bea}
\def \ee {\eea} 

\def \aa  {{\rm a}}
\def \ep   {\epsilon} 
\def \ad {AdS_2 \times S^2} 

\def \be {\bea}
\def \ee {\eea}
\def \adt {$AdS_2 \times S^2$\ }
\def \cc {{\rm c}} 

\def \KK {{\rm K}} 

\def \lm {v} 
\def \tK {\td K} 
\def \mm {{\rm m}}


\tableofcontents

\section{Introduction  and summary}


BPS  Wilson loops (WL) in higher representations in $\cal N$=4 super YM theory 
 admit  a dual  description 
in terms of D-branes \ci{Drukker:2005kx,Gomis:2006sb,Hartnoll:2006is, Okuyama:2006jc, Yamaguchi:2007ps,Drukker:2006zk}.
In particular, in~\cite{Gomis:2006sb} it was shown that WL in the symmetric representation of $SU(n)$ can be described 
in terms of D3-branes, with the  number $k$ of boxes in the Young tableau  playing 
the role of an extra  parameter, in addition  to $N$ and the 't Hooft coupling $\l$. 
For  large $N$    and  $k \ll N$  the dual   description
is given in terms of $k$    coincident strings (same as for multiply wrapped WL) 
  while for $k \sim N \gg 1$ with $k/N$=fixed 
it  is  expected to be in terms of  a probe D3-brane in  \adss space.\foot{For $k \sim N^2 \gg 1$   the probe approximation is no longer valid  as 
 one cannot ignore  back reaction \ci{Lin:2004nb}  of D-branes on geometry \ci{Yamaguchi:2006te,Lunin:2006xr}.}

\iffa
There has been progress on circular Wilson loops in the fundamental representation 
(see also [] for correlation functions of circular Wilson loops and local operators)\\
It is interesting to expend these results for higher representations \\
Wilson loops can be described by  $D$-branes intersection along a line.\\
Gomis and Passerini showed in~\cite{Gomis:2006sb} that Wilson loops in the symmetric representation can be 
described by  intersecting D3-branes and in the antisymmetric representation by D3-branes
intersecting with $D5$-branes. At strong coupling we replace one stack of D3-branes with the $AdS_5 \times S^5$ 
geometry and the effect of remaining branes corresponds to back reaction deforming $AdS_5 \times S^5$. \\
The result of back reaction is the bubbling geometry\cite{Lin:2004nb} and in the case of Wilson loops it was
 constructed in~\cite{Lunin:2006xr}\\
It is hard to use in it practice even for the semiclassical calculation~\cite{Okuda:2008px} \\
Interesting to consider the probe brane limit where we ignore back reaction \\
\fi

The classical probe D3-brane   solution  representing  large $k$  
 circular WL  found  in \cite{Drukker:2005kx}
applies in the limit  of large   $N$ and large $\l$     with 
\be
\k\equiv \frac{ k \sqrt{\l}}{4 N} = {\rm fixed} \ . 
\label{1.1}
\ee
The WL expectation  can   be written as 
\be   \la{2} 
\langle W \rangle = e^{-\G} \ , \ \ \ \ \   \G= \G_0 + \G_1 + ... \ , \ee
where $\G_0$   is the D3-brane probe action evaluated on the classical    solution 
   \cite{Drukker:2005kx}   
\be \la{1} 
\G_0 =  N \F_0 \ , \ \ \ \ \ \ \ \ \   \F_0 =-   2 \big( \k \sqrt{1+\kappa^2} +{\rm Arcsinh}\, \k\big)  \ ,  \ee
and $\G_1$   stands for the first subleading $1/N$ correction. 
For $k \ll N$   one gets  $\G_0= - k \sql$  which is the result for $k$-wrapped circular string WL \ci{ber,dgo}. 
$\G_0$   in \rf{1} was found  to match the corresponding  gauge theory matrix model results   \cite{Drukker:2005kx,Hartnoll:2006is}:
 at this leading order the matrix model  expression  \ci{Erickson:2000af,Drukker:2000rr,Pestun:2007rz}    
 for  circular WL  in  the fundamental representation    extended   
   to $k$-fundamental (or multiply wound)  WL
 case  \cite{Drukker:2005kx}
is the same as  the  WL   corresponding to the symmetric representation   \cite{Hartnoll:2006is, Okuyama:2006jc}.

 As for the   subleading correction $\G_1$, it  appears to be non-trivial to extract it from the general result
  for the rank $k$ symmetric representation WL  given  in \cite{Fiol:2013hna}  but it can be
   readily   found  for the $k$-fundamental  WL case.
 Starting   with the matrix model solution of 
 \cite{Drukker:2000rr}  and replacing $\sql \to k \sql$ one gets  \cite{Drukker:2005kx}\foot{Given the 
 matrix model  expression  $\langle N^{-1} \tr e^{ M} \rangle =  Z^{-1} 
  \int [dM]  \  N^{-1} \tr e^{ M}  \, \exp ( - 2 N \l^{-1} \tr M^2)$ 
 for the WL expectation value in the  fundamental representation,   to find the 
 expectation value in the tensor product of $k$ fundamental representations 
 $\langle N^{-1} \tr e^{k M} \rangle$  one is to replace $k M \to M$, i.e.  $\l \to k^2 \l$. 
 The result  then depends on $N, k, \sql \ $  only through $N$ and  $k\sql $, or $N$ and $\k$.} 
\be 
\langle W_{\Box^k} \rangle = \ e^{- \G_{\Box^k}}  =  {N^{-1}}  e^{2 N \k^2} \, L^{1}_{N-1} (-4 N \kappa^2)\,. 
\label{b2}
\ee
Here $L^m_n$ is the Laguerre polynomial, so that $\G_{\Box^k}(N,\k) $   satisfies 
 \cite{Drukker:2005kx} 
\be  \G_{\Box^k} \equiv N \F \ , \qquad \qquad  
(\pt_{\k} \F)^2 -   N^{-1}  \big( \pt^2_{\k} {\F} + 3\k^{-1}  \pt_{\k} {\F}\big) - 16  (1+ \k^2)=0\, . 
\label{b3}
\ee
For large $N$ and fixed $\k$   one finds
\be \la{3} 
  \F= \F_0(\k)  + N^{-1}  \F_1 (\k)  +   O(N^{-2})  \ , \ee
 where $\F_0 $   is  the same as in \rf{1}
and   the next correction  
  is    \ci{Kawamoto:2008gp}
 \be \la{30} 
 ( \G_1)_{\Box^k} = \F_1= \ha  \ln\big( \k^3 \sqrt{1+\k^2}\, \big) \ . \ee
   In contrast to   $\langle W_{\Box^k}\rangle$  which depends only on 
   $N$ and $k \sql$, 
  the rank $k$ symmetric representation   expression  $\langle W_{\smm_k}\rangle$  
   is, in general,  a non-trivial function of  the three parameters $N, \sql$ and $k$ \ci{Fiol:2013hna}. 
   The  analog of the 
   above  expansion  \rf{3}  will  be  to take   $N$  large 
   first   for fixed $\k$ and $\sql$ and then expand  in large $\sql$ 
   at each order in $1/N$. 
   
There is no a priori reason why the  leading large $\sql$ term in the  first  $1/N$ correction  in  $\langle W_{\smm_k}\rangle$ 
should  be the same as \rf{30}. 
However, it is easy to see that  in the limit $1 \ll k \ll N$, that is in the limit of small $\kappa$,      
the logarithms  of the dimensions  of the two  representations 
$d_{\Box^k} = N^k$ and $d_{\smm_k}=\frac{(N+k-1)!}{k! (N-1)!}$ are the same to leading order. 
Hence, using~\eqref{30} one may  expect that    
  \be \la{33} 
  \k\ll1 \ : \ \ \quad  \ \ \qquad 
 ( \G_1)_{\Box^k}  \approx   ( \G_1)_{\smm_k}   \approx          \ha  \ln  \k^3  \ . \ee
Our aim  below    will be  to see  how   this result can be reproduced  in the dual D3-brane description  \cite{Drukker:2005kx}
of the symmetric represention  WL.\footnote{One can expect that the same conclusion should also hold in the case of the totally antisymmetric 
representation. It indeed holds for the leading semiclassical result at large $\lambda$ obtained in~\cite{Hartnoll:2006is, Okuyama:2006jc}:
$\langle W_{A_k} \rangle ={\rm exp} (2 N \sqrt{\lambda} \sin^3 \theta/3 \pi)$, where $\theta -\frac{1}{2}\sin 2 \theta =\frac{\pi k}{N}$. 
Taking the limit $k \ll N$ we find that $\langle W_{A_k} \rangle \to {\rm exp} ( k \sqrt{\lambda})$ which coincides with the small $\kappa$ 
limit of~\eqref{2}, \eqref{1}.}

One  natural suggestion is  that 
to go beyond the leading large $N$   result  \rf{1} 
one should include the contribution of one-loop  fluctuations of D3-brane fields  near the  classical solution 
 of  \cite{Drukker:2005kx}  (ignoring   $\a'$ corrections to the  D3-brane probe  action and  also 
 contributions of loops of massive string  modes as they should be suppressed at leading order in large $\sql$). 
 Since the D3-brane probe tension in \adss  is effectively 
 proportional to $N$,   the 
 1-loop correction  $\G_1$  to the  D3-brane  effective  action \rf{2} 
 will be a function of $\k$ only. 
 The path integral  for the D3-brane  probe action  is not well-defined  in general 
 (the DBI  action is non-renormalizable), but  the semiclassical   correction
  may still make sense 
 in the expansion near a   BPS  solution  as it 
 corresponds to taking into account the contributions of massless   open string modes  only. 
  While the  resulting expression   for $\G_1$ will   be UV divergent,   indicating the need to account  for 
 a  proper embedding of this calculation into string theory, 
 one may still expect that  the finite $\k$-dependence
  is  correctly captured (to leading order in large  $\sql$) 
  by the massless   mode contributions only. 
  Indeed, the string theory cutoff should be independent of the background parameter $\k$, 
  and the massive string mode contributions   should be suppressed   by extra  powers of the inverse  of string tension $ \sql $.

 The    investigation  of  the semiclassical quantization of the probe D3-brane action near the    solution  \cite{Drukker:2005kx}
 representing  the  circular WL  was  initiated  in \ci{Faraggi:2011bb,Faraggi:2011ge}  which we will  build on. 
  The   action  for the  quadratic  fluctuations   is essentially that of an  Abelian ${\cal N}=4$ vector 
  multiplet in $AdS_2 \times S^2$ background  with    $\k$-dependent  radii.  
    Computing the corresponding   one-loop correction to the classical  value of D3-brane action  we will find that 
  \be \la{03}
  \G_{\rm {1}} =  \ha  \ln (L^2 \Lambda^2) +   \G_{\rm {1\, fin}} (\k)  +   C_1     \ , 
  \qquad \qquad     \G_{\rm {1\, fin}} (\k) =   \ha \ln { \k^3\ov  \sqrt{1 + \k^2} } \ , \ee
  where $ \L$ is UV cutoff, $L$ is  \adss radius  and $C_1$   is a  numerical ($\k$-independent)
   constant  given by the   sum 
  of $\zeta'(0)$ terms  for the corresponding  quadratic fluctuation operators in $AdS_2 \times S^2$  geometry.

  \def \vee  {\varepsilon} 
  
  We   conclude that    the $\k$-dependent part of the one-loop D3-brane result  \rf{03}   
  differs in general  from   the $k$-fundamental  WL expression \rf{30},  but the two do   coincide   in the 
  small $\k$ limit, in agreement with \rf{33}. 
  We conjecture   that $ \G_{\rm {1\, fin}} $  in \rf{03} 
  should match the symmetric representation expression $( \G_1)_{\smm_k} $   for any finite $\k$. 
  
 Below in section  2.1   we shall first review the classical D3-brane solution \cite{Drukker:2005kx} representing  circular WL 
in the symmetric representation at strong coupling  and  then  in section 2.2
describe   the  quadratic fluctuation part of the D3-brane probe action expanded near
 it following  \ci{Faraggi:2011bb,Faraggi:2011ge}. 
 The  $\k$-parameter dependent part $  \G_{\rm {1\, fin}} (\k) $ 
     of the corresponding one-loop effective action \rf{03}  will be  found in section 2.3. 
     
     In Appendix A we shall  compute  the numerical  part $C_1$ of the effective action \rf{03} 
     using heat  kernel methods \ci{cam,camp,ban,gul}  in \adt background. 
     In Appendix B we shall review  the structure of the one-loop  correction \ci{dgt}  to circular WL in  \adss   string theory 
     (corresponding to  the  limit of $N=\infty$ with $\l \gg 1$  for fixed $k$)   and 
     then   rederive  the result of \ci{kt} for  its finite  part 
     for $k=1$ by completing  the computation originally  presented  in \ci{dgt}  based on   $AdS_2$ heat kernel  expressions.



\section{One-loop    correction  near  the D3-brane
   solution } 

\subsection{Classical solution}

Let us  first review the   classical solution  \cite{Drukker:2005kx}  for D3-brane probe in \adss    
that should be  describing  \cite{Gomis:2006sb}
 the expectation value of a BPS  circular  Wilson loop in  the rank $k$  symmetric representation of $SU(N)$. 
The D3-brane is  localized in $AdS_5$  with    coordinates  chosen as 
\bea ds^2= \frac{L^2}{\sin^2 \eta} \big[ d\eta^2 + \cos^2 \eta \, d \psi^2 + d \r^2 +\sinh^2 \r \, 
(d \q^2 +\sin^2 \q\,  d \phi^2)\big]   \,. 
\label{2.3}
\eea
Here  $L$   is the \adss radius, \  $L^2 =  (4 \pi g_s N)^{1/2}\,  \a'  \equiv \sql \a' $. 
The worldvolume of the D3 brane will be parametrized by $\s^{\a}= (\psi, \r, \q, \phi)$. 
The bosonic part of the   D3-brane  probe action  in  \adss   is 
\bea 
S=
 T_{D_3}\Big(  \int d^4 \s \ e^{- \Phi} \, 
\sqrt{{\rm det} (g+ 2 \pi \alpha' F)}\,
 -   \int C_4 \ \Big) \ ,  \qquad T_{D_3}=\frac{1}{(2 \pi)^3 g_s \a'^2}= \frac{N}{2 \pi L^4} \ .
\label{2.4}
\eea
%
 The relevant solution
 ending on a circle parametrized by $\psi$  at the boundary  $\eta=0$   is  \cite{Drukker:2005kx}\footnote{The 
 gauge field  has  an  extra factor of $i$ 
 in Euclidean space (it becomes   a real  electric field in Minkowski signature). 
 The momentum  conjugate to $A_\psi$    component  of vector potential 
 is set  equal to  $k$   which  on the dual gauge theory side 
 is identified   with   the rank  of the symmetric representation.}
\be \sin \eta ={\kappa}^{-1}\, \sinh \r \,,\ \ \ \ \  \qquad 2\pi \a' \bar F_{\psi \r}= \frac{i  L^2}{ \sinh^2 \r}\,, 
\ \ \ \ \ \ \   \k=\frac{k \sqrt{\lambda}}{4 N}= { k L^2 \ov 4 N} \,. 
\label{2.10}
\ee
The induced metric on the D3-brane is then that of  \adt
 \be
 &&d{\bar s}^2=   a_{AdS_2}^2(d \xi^2 + \sinh^2 \xi\,  d \psi^2) +   a^2_{S^2} (d \q^2 + \sin^2 \q\,  d \phi^2)\,, 
\label{212}  \\
&& a_{AdS_2}= L \sqrt{1+\k^2}\ , \qquad  a_{S^2} =L \k \ , \qquad     \sinh \xi = (1 + \k^2)^{-1/2}  \cot \eta
\ . \la{21} \ee
Substituting the solution~\rf{2.10} into the action \rf{2.4} 
(and taking into account  boundary terms) 
one finds     that  the latter  is equal to $\G_0$ in \rf{1}   \cite{Drukker:2005kx}.

 Let us note that  the BPS solution  corresponding to the ($k$-fold) infinite straight line WL 
 can be found from~\eqref{2.10} by taking $\eta$ and $\r$ small
and replacing $\psi$ with a non-compact coordinate $x$. 
Both  circle and  straight line  solutions 
can be described in a unified manner in the coordinate system where the $AdS_5$ metric is \cite{Faraggi:2011bb}
\be 
ds^2 = L^2\big(d u^2 + \cosh^2 u\, ds_{AdS_2}^2 + \sinh^2u\, ds^2_{S^2}\big)\,. 
\label{2.18}
\ee
   Choosing    $ds^2_{AdS_2}$    in Poincare coordinates, i.e.  
   $ ds^2_{\widetilde{AdS}_2}= {r}^{-2} ( d{\rm x} ^2 + d {r}^2)$,  corresponds to  the straight line  case 
   and  choosing it  in global  coordinates  $ ds^2_{AdS_2}=d\chi^2 +\sinh^2 \chi\, d \psi^2$ -- to the circle case. 
   Then in both cases   the solution is  simply  $u=u_\k  ,\   \sinh u_\k =\k $, so that   
the induced metric  (cf.  \rf{212})  and the  electric  gauge field background are  
\be
 {\bar g}_{\a \b} d \s^{\a} d {\s^b}=  L^2 \big( \cosh^2u_\k\,ds^2_{AdS_2} + \sinh^2u_\k\, ds^2_{S^2}\big)\,, \   
\ \ \ \ \ \ \ 
  \sinh u_\k = \k  \,, \     
\label{2.25}\\
 2 \pi \a' \bar F = i L^2 \cosh u_\k\,  e^0 \wedge e^1\ , \qquad \qquad \qquad \qquad \qquad \qquad \la{bac}
 \ee
 where $(e^0,e^1)$ is  the vielbein  of the Euclidean $AdS_2$. 
 
 Let us recall   that   the euclidean $AdS_2$  space in Poincare coordinates  with  boundary $R$ 
 (which we shall denote as ${\widetilde{AdS}_2}$)   is 
not   globally equivalent to $H_2= AdS_2$   in global coordinates  with  boundary $S^1$.  
In particular, their  regularized  and renormalized volumes are different 
\be  V_{{\widetilde{AdS}_2} }= { Ta\ov \vee} \to\  0 \ , \ \ \ \ \ \ \ \ \ \ \ \ \ 
 V_{{{AdS}_2} }=2 \pi  a^2( { 1 \ov \vee} - 1)   \to \  - 2 \pi a^2 \ . \la{voo}\ee
 Here $a$ is the  radius,  $T$ is the length of the boundary of ${{\widetilde{AdS}_2} }$
  and  $\vee$ is a radial  IR cutoff.
  
 In what follows we shall always assume, as by now  standard, 
  that all power IR  divergences are ignored (being  cancelled by proper boundary terms), so that   volumes should  be replaced by their renormalized
   values. 
 As a  result, in the straight  line WL  case  the classical  D3-brane probe action vanishes  \cite{Drukker:2005kx}
 (consistent with  $\langle W \rangle =1$ in this 1/2 BPS case), while in the   circular WL case  one  finds 
 a non-trivial expression \rf{1}. This is, of course,  parallel to  what happens in the  corresponding 
  string  theory description of these WL's   at finite $k$ \ci{dgo,dgt,sz,kt}. 
   Same  will also apply     at the  loop level.

%


\subsection{Quadratic fluctuation  action}

Our aim is to compute the first quantum  correction to the D3-brane  effective action  expanded  near the above classical 
solution. As the   D3-brane action  \rf{2.4}   scales as $N$ (for fixed $L$),  the  one-loop 
 correction  corresponds to the first subleading order in 
$1/N$ expansion at fixed $\k$. 

The derivation of the quadratic   fluctuation action was described in detail in \ci{Faraggi:2011bb,Faraggi:2011ge}
which we follow here. In a static-like  gauge the remaining 6 bosonic  scalar   fluctuations are 
$\Phi^I= L\, ( \d u,  \d \q^a)$  where $u= u_\k + T_{D_3}^{-1/2}\d u$  and $ \d \q^a$ are 5 
 fluctuations  in $S^5$   directions. There are also gauge field fluctuations  defined by 
 $2\pi\a' F_{\a\b} =
 2\pi\a' \bar  F_{\a\b} +  T_{D_3}^{-1/2}   f_{\a\b}$. 
 The quadratic fluctuation action is the sum of the  bosonic and fermionic parts 
  $\td S=\td S_B  + \td S_F$
 and   is  esentially  the same as the  action  of an Abelian ${\cal N}=4$ supersymmetric vector multiplet 
in curved  $AdS_2 \times S^2$  background:\foot{The simplicity   of the fluctuation action is due 
to the residual supersymmetry of the background  BPS solution.}
 %
\bea 
\td S_B &= & \int d^4 \s \sqrt{\bar{M}} \, \te  \Big(\frac{1}{2} G^{\a \b} \pt_{\a} \Phi^I  \pt_{\b} \Phi^I +\frac{1}{4} G^{\a \b} G^{\g \d}f_{\a \g}  f_{\b \d}\Big)
\nonumber \\
& = &   \cc  \int d^4 \s \sqrt{G}\, \te \Big(\frac{1}{2} G^{\a \b} \pt_{\a} \Phi^I  \pt_{\b} \Phi^I +\frac{1}{4} G^{\a \b} G^{\g \d}f_{\a \g}  f_{\b \d}\Big)\, , 
\label{2.31}\\
\td S_F &=&  \int d^4 \s \sqrt{\bar M} \  G^{\a\b} \Theta ( i \Gamma_{\a}  \nabla_{\b}) \Theta= 
 \cc  \int d^4 \s \sqrt{G} \ \Theta (i \Gamma^{\a}  \nabla_{\a}) \Theta\,.
 \label{2.34}
 \eea
Here  ${\bar M}$
is the determinant of the  matrix 
\be 
&&{\bar M}_{\a \b}= {\bar g}_{\a \b}+ 2 \pi \a' {\bar F}_{\a \b}\,, 
\label{2.31.1}\\
&& \sqrt{\bar M} =  \cc\,  \sqrt{G}\, , \ \ \ \ \ \ \ \ \ \    \cc \equiv \coth u_k = {\sqrt{1 + \k^2} \ov \k}   \ ,  \label{222}
\ee
where ${\bar g}_{\a \b}$ and ${\bar F}_{\a \b}$ are the induced metric and the classical solution for the gauge field.
$G$   is  the determinant of the ``open string"   metric  
\be 
G_{\a \b} ={\bar g}_{\a \b} + (2 \pi \a')^2 {\bar g}^{\g \d} {\bar F}_{\a \g} {\bar F}_{\b \d} \,, 
\label{2.32}\ee
which is  the inverse of the symmetric part of ${\bar M}_{\a \b}$, 
i.e. ${\bar M}^{(\a \b)} = G^{\a\b}$ (see, e.g.,   \ci{Seiberg:1999vs}).
In  the present case  $G_{\a\b}$  turns  out to be that of the \adt space with {\it equal} radii (cf. \rf{2.25})
\be  
G_{\a \b} d\s^{\a}  d\s^{\b} =a^2  \big(ds^2_{AdS_2} + ds^2_{S^2}\big)\,, \ \ \ \ \ \ \ \ \ \ \    a= L \k \ . 
\label{2.33}
\ee
The spinor covariant derivative $\nabla_\a$ in \rf{2.34}   is defined with respect to $G_{\a\b}$. 

The action \rf{2.31},\rf{2.34}    is different  from the action  for a  massless  $\cal N$=4 vector  multiplet 
in \adt geometry  only  by the overall  constant factor  $\cc $.
The presence of this  factor  which originates 
from   $ \sqrt{\bar M}  $ in  \rf{222}  
implies  that   the covariant  measures  for the  
fluctuations  in question (which are essentially  massless   open-string modes)  
should   be  defined  with respect to 
 ${\bar M}_{\a \b}$ as\foot{Here  $a_\a$   is  a vector potential,  $f_{\a\b}= \del_\a a_\b - \del_\b a_\a$.}
%
\bea &&
|| \Phi||^2 =\int d^4 \s \sqrt{\bar M} \ \Phi^I  \Phi^I = \cc  \int d^4 \s \sqrt{ G} \, \Phi^I  \Phi^I  \,,\la{23}  \\
&&  
|| a ||^2 =\int d^4 \s \sqrt{\bar M} \  M^{\a \b}\  a_{\a}  a_{\b}= \cc  \int d^4 \s \sqrt{ G} \,  G^{\a \b}\, a_{\a}  a_{\b}   \,, 
\la{22}  \\
&& 
|| \Theta||^2 =\int d^4 \s \sqrt{\bar M} \ \Theta \cdot \Theta =  \cc  \int d^4 \s \sqrt{ G} \, \Theta \cdot \Theta  \, .
\label{24}
\eea

\subsection{One-loop correction to  D3-brane effective action} 

In general, the one-loop   effective action of   the $\N=4$   vector multiplet theory in curved   background is given by 
the standard sum of the gauge field, 6  real scalar and 4 Majorana fermion  contributions
\be 
 &&\ \ \ \ \ \ \ \ \ \ \ \ \  \ \ \ \ \  \G_1= \G_g  + 6 \G_s     + 4 \G_f \ ,\la{311} \\
 && \te  \G_g = \ha \ln \det (- G_{\a\b} \nabla^2 +  R_{\a\b} ) -  \ln \det (- \nabla^2)  \ ,\ \ \ \ \
  \G_s = \ha \ln \det (- \nabla^2 + {1\ov 6} R) \ , \la{31}  \ \ \ \ \ \ \ \ 
 \\
&&\te 
\G_f  =- \ha  \ln \det ( i \G^\a  \nabla_\a ) =- {1 \ov 4} \ln \det ( -  \nabla^2 + {1\ov 4} R )  \ , \la{32}
\ee
where the second term in $\G_g$ corresponds to the contribution of the ghosts. 
In the present case of  equal-radii   conformally-flat  \adt   geometry   \rf{2.33}    the Ricci scalar and the Weyl tensor of $G_{\a\b}$
vanish.\foot{Thus    the scalars $\Phi^I$ in \rf{2.31}  are, in fact,  conformally  coupled.}
The  above  determinants   are assumed    to be  defined with respect to the  measures in \rf{23}--\rf{24}. 

Since  \adt is a homogeneous  space, $\G_1$ will be proportional  to  its volume. 
As already mentioned above,  we shall assume that  all power IR divergences should be ignored, 
i.e. the volume of \adt   space with boundary $S^1 \times S^2$ should be  replaced by its 
renormalized value 
\be 
 V_{\ad}= ( 4 \pi a^2) \times (- 2 \pi a^2) = - 8 \pi^2 a^4 \ . \la{v} 
\ee
In the  straight-line solution case  the  renormalized  volume
  will vanish (see \rf{voo}),   $V_{\widetilde{AdS}_2\times S^2}=0$,  
  and thus, as in the corresponding  string-theory  description 
  \ci{dgt,kt}, the one- (and higher-) loop corrections   will vanish too.

 In the circular solution case $\G_1$ will be non-zero and   will   contain  three  terms (cf. \rf{03}): 
 
 \noindent  (i) a 
  UV divergent  (conformal anomaly  related) term  $\G_\infty$  proportional to the logarithm of the product of (e.g.,  proper-time
  $\ep=\L^{-2}\to 0$) 
  UV cutoff    and  the radius $a$ in \rf{2.33};  
  
  \noindent 
 (ii)  a finite  contribution $\G_m  \sim \ln \cc $     of the non-trivial measure;  
 
 \noindent 
 (iii)  a non-trivial  numerical constant  $C_1$ 
  (given by the sum of  finite parts of  logs of normalized determinants)
   that can be found  using   standard  heat kernel techniques
  \ci{camp}  (as   discussed for  4d scalars, vectors and spinors in  \adt   background in a   different  context in \ci{ban,gul}).

  We will  present  the   computation of $C_1$  in Appendix A while here will
  concentrate on the first two contributions. 
   Explicitly, we  have 
  \be \la{se}
   \G_{\infty}  = - \ha B_4  \ln  ( a^2 \L^2) \ , \ \ \ \ \ \ \ \ \ \ \ \    B_4 = \te {1\ov (4\pi)^2} \int d^4\s\, \sqrt{G}\,  b_4 \ , 
   \ee
where $b_4$  is the local Seeley coefficient (see, e.g., \ci{duff,Vassilevich:2003xt}). 
In the present \adt case 
\be 
&& b_4 = - \aa\,  R^*R^* \ , \ \ \ \  \ \ \  R^*R^*= - 2 R_{\a\b}^2 =- 8 a^{-4}  \ , \ \ \ \ \ \la{bb} \\
&&  \te \aa= \aa_g + 6 \aa_s   +  4 \aa_f = {31\ov 180} + 6 \times { 1\ov 360}  +  4 \times  { 11 \ov 720}  = {1 \ov 4} \ , \la{aa}\\
&& B_4 = \te {1\ov (4\pi)^2} V_{\ad} \, b_4 =  -1    
\ ,  \la{v1}\ \ \ \ 
\ee
where we used \rf{v}. 
Thus \rf{se} becomes 
\be 
 \G_{\infty}  
 = \ha  \ln (  a^2 \L^2)  \ . \la{cc} 
\ee
The  regularized measure contribution is also   controlled   by the same Seeley  coefficient $B_4$
appearing in the expansion of $\tr 1  \to \tr e^{-\ep \Delta}$  ($\Delta$ is the corresponding 2nd order operator),\foot{Note that the  squared  fermionic operator 
is assumed to be  defined  with respect to the same   measure \rf{24}. Then all power divergences cancel,  consistent  with supersymmetry of this model.
In general, given a spectral problem  for an  operator $\cO$ with respect to the  measure $\m$,
 i.e. $ \cO f_n = \l_n f_n, \ \int \mu\, f_n^* f_m = \delta_{nm}$,  the path integral with the action $S= \int \mu\, \phi^* \cO \phi $
 is   expressed in terms of $\det \cO= \prod_n \l_n$  (one sets $\phi= \sum_n c_n f_n $ and integrates over $c_n$). 
Then $\det \, \cO^2= \prod_n \l^2_n $   corresponds to path integral with the action $S= \int \mu\, \phi^* \cO^2 \phi $ and 
the constant  measure  factor  dependence is   controlled by the Seeley coefficient of $\cO^2$
(see, e.g., Appendix A in \ci{dgt}).
} 
i.e.  
\be 
 \G_{m}  =  \ha B_4  \ln \cc  = - \ha \ln \cc   \ . \la{dd} 
\ee
Since  the UV cutoff   should  not depend on the parameter $\k$ of the background solution
(it should  be, in fact,  proportional to $\a'^{-1}$  in a consistent string theory embedding) 
we can then determine  the $\k$-dependent  part  of $\G_1$ by combining \rf{cc} and \rf{dd}
and using  the definitions of $\cc$ \rf{222} and $a$ \rf{2.33}:
\be 
\G_{\infty} + \G_m =  \ha  \ln (  L^2  \L^2)    +   \G_{1\,\rm fin}(\k) \ , \ \ \ \ \ \ \ \ \ \ \ \ 
 \G_{1\,\rm fin}(\k)= \ha \ln {\k^3\ov \sqrt{1 + \k^2}} \ .  \la{ee} \ee
 We thus  find  the expression   in  \rf{03}   announced  in the Introduction.


\section*{Acknowledgements}
We  would like to thank  N. Drukker, B. Fiol, M. Kruczenski, Yu. Makeenko  and A. Tirziu  for very 
useful  discussions.
The work of E.I.B. was supported by the ARC Future Fellowship FT120100466.
The   work of A.A.T. was supported by the ERC Advanced grant No.290456
and also by the STFC grant ST/J000353/1.
This work was  also supported in part by the ARC Discovery project DP140103925.
E.I.B. would  like to thank Theory Group at Imperial College  where the part of the work was done
for warm hospitality.

\def \ss {{\rm s}}  \def \ff {{\rm f}} \def \gv {{\rm g}} 
\def \vv {{\rm v} }
\def \ddd {d} 
\appendix
\section{Finite part of one-loop effective action of  $\N=4$   vector multiplet on \adt}
\def\theequation{A.\arabic{equation}}
\setcounter{equation}{0}

The contributions of   determinants of 2nd order operators  in \rf{31}, \rf{32} 
can be expressed  in terms of  the trace of the corresponding   heat kernel in the standard way.
In the present  case of   \adt  space \rf{2.33}   
\be \la{a1} 
\ha \ln \det \Delta = - \ha \int_\ep^\infty { dt \ov t} \  \KK(t) \ , \ \ \ \ \ \ \ \ \ \ \ 
\KK(t) =   V_{{\ad }} \  K(t)\  . 
\ee
Here $\KK(t) = \sum_n e^{-\l_n t}$    and for a  homogeneous space 
 $K(t) \equiv K(x,x;t) = \sum_n e^{-\l_n t} f_n^* (x)  f_n (x)$   does not depend on 
 the point $x$ ($\{f_n (x)\} $ is  a   set of normalized eigenfunctions 
 of $\Delta$ with eigenvalues $\l_n$). 
The  renormalized   expression for the \adt  volume  factor was given  in \rf{v}.

In general, the  finite   part of  \rf{a1} is naturally expressed as $-\ha \z'(0)$  in terms of  the corresponding $\z$-function 
\be 
&&\z (z)= \frac{1}{\G (z)} \int_0^{\infty} dt\,  t^{z-1}\,  \KK(t) =  \frac{1}{\G (z)} V_{\ad} \int_0^{\infty} dt\,  t^{z-1}\,  K(t)        \,,
\label{4.1}\\
&& \la{aa1} 
\ha \ln \det \Delta =  - \ha \z(0) \ln (1/ \bar\ep)    - \ha \z'(0)   \ . \la{z}\ee
Here $\bar \ep =a^{-2}\m \ep $ is  the  combination   of the UV cutoff $\ep= \L^{-2}$, geometrical scale $a$ and 
a possible   measure factor $\m$ (corresponding to $\Delta \to \m \Delta$). If we formally include in $\z(0)$  
 the contribution of potential  zero modes (by IR-regularizing them with a small mass term)
   its  expression    will   match  the  value  of  the Seeley coefficient  $B_4$  in \rf{se}, \rf{dd}.

For a  product space   like \adt  the   scalar and spinor heat   kernel    factorizes. 
For completeness, let us  first quote  the  ``untraced''   expressions for  heat kernels $K(x,x'; t) $ 
  of  massless scalar  Laplacians on $S^2$ and $AdS_2$  (see, e.g., \ci{cam,cha,jon}). 
  They depend on coordinates of two  points $x,x'$ only through  the corresponding 
  geodesic distance $\ddd(x,x')$. 
For  $S^2$   with  unit-radius metric $ds^2 = d \q^2 + \sin^2 \q\,  d \phi^2$, \  $x=(\q,\phi)$,  
 the  geodesic   distance  is given by $\cos \ddd(x,x')
= \cos \q\,  \cos \q'+ \sin \q\,  \sin \q'\,  \cos(\phi-\phi')$ and\foot{The derivation starts with $K$ written as a sum of product of 
 normalized    eigenfuctions 
 (spherical harmonics)    and  uses  the  summation formula    for the associate Legendre polynomials: 
 $ P_\ell (\cos  \ddd(x,x') ) = P_\ell( \cos \q)  P_\ell( \cos \q')  + 2 \sum_{m=1}^\ell {(\ell-m)!\ov (\ell + m)!} P^m_\ell( \cos \q)  
 P^m_\ell( \cos \q') \cos (\phi-\phi')$,  with  $ \ P^{-m}_\ell( \cos \q) = (-1)^m {(\ell-m)!\ov (\ell + m)!} P^m_\ell( \cos \q) $.} 
\be 
K_{S^2}(x,x';t) ={\te  { 1 \ov 4 \pi}} \sum_{\ell =0}^\infty (2 \ell+1)   P_\ell (\cos  \ddd(x,x') ) \,  e^{- t \ell (\ell + 1) } 
\ . \la{xs} \ee
For  $AdS_2$   with metric $d \xi^2 + \sinh^2 \xi\,  d \psi^2$,\    $x=(\xi,\psi)$,  
 the  geodesic   distance  is  
 $\cosh \ddd(x,x')
= \cosh \xi\,  \cosh \xi'-   \sinh \xi\,  \sinh \xi'\,  \cos(\psi-\psi')$  and \foot{The expressions 
\rf{xs} and \rf{xa}   are formally related  by an analytic continuation 
and use of   the  relation 
$\sum_\ell   f(\ell)  = {1\ov 2 i} \int_C  dz\,  \cot (\pi z)\,  f(z)$   where $C$ encirles the real axis. 
Explicitly, one is to set   $\ell \to z = -\ha  +  i v$,    continue the angles 
$\q \to i \xi, \ \phi \to \psi$ so that $\ddd_{S^2}(x,x')  \to \ddd_{AdS_2}(x,x')$, and finally 
 reverse  the overall sign of the metric (or  restore  the radius   factor   and set $a\to i a$).
The latter   corresponds to $t \to - t$ in \rf{xs}, as  required to match \rf{xa}.} 
\be 
K_{AdS_2} (x,x';t) ={\te  { 1 \ov 2 \pi}}\int^\infty_0  dv \,  v \, \tanh ( \pi v)\, P_{- {1\ov 2} + i v } (\cosh \ddd(x,x'))\ e^{ - t ( v^2 + {1\ov 4})} 
\ .   \la{xa} \ee
Here   $P_q (y) $ is  the  Legendre function  that becomes equal to 1 at coincident points  when 
$\ddd(x,x)=0$, i.e.  $P_q(1)=1$.  Let us note that  $K_{AdS_2} (x,x';t) $   admits also an alternative representation \ci{cha}: 
\be 
K_{AdS_2} (x,x';t) ={\te  { \sqrt 2  \ov (4 \pi t)^{3/2} }}\int^\infty_{\ddd (x,x')} 
  du \,  { u   \ov  \sqrt{ \cosh u -  \cosh \ddd(x,x')   }}   \ e^{ - {1\ov 4} ( { u^2\ov t}  + t )} 
\ .   \la{xaa} \ee

The expressions for   $K(t)=K(x,x;t)$  for the massless scalar, vector and fermion operators in \rf{31},\rf{32}  can be found 
in \ci{cam,camp,ban,gul}  and are summarized  below.   For a  real scalar operator on $AdS_2 \times S^2$   with equal radii  
$a$  one has 
\be 
&& K^\ss (t)= K^\ss_{AdS_2} (t)\,  K^\ss_{S^2} (t)\,,\la{a3} \\
&&
K^\ss_{S^2} (t) ={\te \frac{1}{4 \pi a^2}} \sum_{\ell=0}^{\infty} (2 \ell+1) \ e^{- t \ell (\ell+1)a^{-2}}\,,\la{aa3} \\
&& K^\ss_{AdS_2} (t)= {\te \frac{1}{2 \pi a^2}} \int_{0}^{\infty} d \lm \, \lm\, \tanh (\pi \lm) \ e^{-t (\lm^2 +\frac{1}{4})a^{-2}}\,. 
\label{3.9}
\eea
Here $v$  is  a continous   spectral parameter  for  the Euclidean $AdS_2$ case.\foot{Note that there is no scalar 
zero mode on the total \adt  space  despite 
the presence of $\ell=0$ zero  mode on $S^2$.}
From now on we shall set  the radius $a=1$  as   the dependence on it is controlled by the conformal anomaly 
coefficient  $B_4$ and   was already determined in \rf{cc}. 

For  a single  4d Majorana  fermion  contribution one gets
\be 
&&K^\ff (t)= -  \ha K^\ff_{AdS_2} (t)\,  K^\ff_{S^2} (t)\,,\la{a4} \\
&&
K^\ff_{S^2} (t) =- { \te \frac{1}{2 \pi}} \sum_{\ell=0}^{\infty} (2 \ell+2)\ e^{-{ t}  (\ell+1)^2}\,, \la{a5}
\quad \ 
K^\ff_{AdS_2} (t)= -{\te  \frac{1}{ \pi }} \int_{0}^{\infty} d \lm\, \lm\, \coth (\pi \lm)\ e^{-{ t} \lm^2}\,. \ \ \ \ 
\eea
The gauge vector  field   contribution in \rf{31} can be written as  the difference of the  contribution $K^{\vv}$ 
of  the operator $- G_{\a\b} \nabla^2 + R_{\a\b}$ 
and  that of the 2 scalar ghost  operators   \ci{ban} 
\be 
K^\gv = K^\vv  - 2 K^\ss \ , \ \ \ \ \ \ \  K^\vv = K^\vv_{AdS_2} \,  K^\ss_{S^2}  +  K^\vv_{S^2}  \, K^\ss_{AdS_2} \ . \la{a6}
\ee
The  vector heat kernel  on $S^2$  is the same as  that of two scalar ones with 0-mode   subtracted, while 
on $AdS_2$ there is an additional vector mode that  is not related   to the scalar ones, i.e. 
\be 
\te K^\vv_{S^2}= 2 \big( K^\ss_{S^2}- \frac{1}{4 \pi }\big)\,, \ \ \ \ \ \ \ \ \ 
K^\vv_{AdS_2}= 2 K^\ss_{AdS_2} +\frac{1}{2\pi }\ . 
\label{3.24}
\ee
As a result, $K^\gv$  can be written as the contribution of two scalars  plus an extra  term:
\be 
K^\gv = 2 K^\ss + \td K\  ,\ \ \ \ \ \ \ \ \ 
   K^\ss  = K^\ss_{AdS_2}(t)\,  K^\ss_{S^2}(t)\ , \ \ \ \ \ \ \ 
   \td K= \te \frac{1}{2\pi}\big[ K^\ss_{S^2}(t)- K^\ss_{AdS_2}(t)\big]\  . \label{326}
\ee
The total  traced  heat kernel   contribution of  the  fields of  an Abelian ${\cal N}=4$ multiplet  is then 
\be 
K (t) = 8 K^\ss_{AdS_2}(t)\,  K^\ss_{S^2}(t)  +4   K^\ff_{AdS_2}(t)\, K^\ff_{S^2}(t)+ \te \frac{1}{2\pi }\big[ K^\ss_{S^2}(t)- K^\ss_{AdS_2}(t)\big]\,. 
\label{3.27}
\ee
%
From \rf{a3}--\rf{3.9} we get then  for the $\z$-function \rf{4.1} in  the scalar   operator  case (using  the  expression for  
$V_{\ad}$ in \rf{v})
\be 
 \z^\ss (z) =  - \sum_{\ell=0}^{\infty} (2 \ell+1) \int_{0}^{\infty} d \lm\, 
\frac{\lm\, \tanh (\pi \lm) }{ \big[\lm^2 + (\ell +\frac{1}{2})^2\big]^z}\,. 
\label{A2}
\ee
To evaluate this  we  may first split $\tanh (\pi \lm)$ into two terms as 
$
\tanh (\pi \lm)= 1-  {2}({e^{2 \pi \lm}+1})^{-1}$.
Denoting  the corresponding contributions  as  $\z^\ss_1 (z)$ and  $\z^\ss_2 (z)$  we find
\be 
&&\te  \z_1^\ss (0) =-\frac{7}{960}\,, \qquad \qquad  \z_1^{\ss\,  \prime} (0)=
- \frac{1}{960} \big[7 -2 \ln2 + 1680\, \z'_R (-3)\big]\,,  
\label{A.17}
\\ 
\label{A.1.8}
&&\z_2^\ss (z)=  2   \int_{0}^{\infty} d \lm\  
\frac{\lm  }{e^{2 \pi \lm}+1}\, H(\lm,z) \ , \ \ \ \ \ \ \ \ \ 
  H(\lm,z) \equiv  \sum_{\ell=0}^{\infty} \frac{2 \ell+1}{ \big[\lm^2 + (\ell +\frac{1}{2})^2\big]^z}   \,,\\
&& 
\z_2^{\ss\, \prime} (0)=-2 \int_{0}^{\infty} d \lm \ 
\frac{\lm  }{e^{2 \pi \lm}+1}\,  H'(\lm, 0)\,,
\label{A.1.10}
\ee
where prime denotes derivative over $z$. 
Expanding the denominator of $H$ in power series in $\lm$ 
and doing the sum using the  standard Riemann zeta function, i.e. $\sum_{\ell=0}^\infty (  2 \ell + 1 )^{-p} = ( 1 - 2 ^{-p}) \zeta_R (p)$, etc.,  
 gives (see, e.g., \ci{cd,ft,Sakai:1984fg})
 \be 
  H(\lm,z) &=&
\sum_{r=0}^{\infty} (- \lm^2)^r \, 2^{2z + 2 r}  \frac{\G (r+z)}{\G (z) \G (r+1)}  \sum_{\ell=0}^{\infty}
(2 \ell+1)^{-2r -2z+1}
\nonumber \\
&=& 2 \sum_{r=0}^{\infty}  (-\lm^2)^r\,  \frac{\G (r+z)}{\G (z) \G (r+1)}  (2^{2r +2 z-1}-1) \,  \z_R (2r +2 z-1)
\,. 
\label{A1112}
\ee
As a result,\foot{We used the identity  

$
\sum_{r=2}^{\infty}   (-\lm^2)^r \, (   2^{2r-1} -1)   r^{-1} \z_R (2 r -1) =  \lm^2 (  \g_E+ 2 \ln 2)   + 
\ha \int_0^{\lm^2} d x\,  \big[ \psi (\ha + i \sqrt{x} ) + \psi ( \ha - i \sqrt{x})\big] .$
}
\be
 H(\lm,0) &=& \te \frac{1}{12}- \lm^2\,, 
\la{aaa1} \\
 H' (\lm, 0) &= &{\te - {1\ov 6}(1 + \ln 2) }  + 2 \ln A +  \int_0^{\lm^2} d x \Big[ \psi \big(\ha  +i \sqrt{x} \big) + \psi \big(\ha - i \sqrt{x}\big) \Big]
\nonumber \\
 &=&\te  -\frac{1}{6} -\ln(2 \pi) -4 \ln A +2  {\cal H} (\lm)\,, \\
\label{A.1.16}
 {\cal H}(\lm)&=&  i \lm\,    \ln { \Gamma \big(\frac{1}{2}- i \lm\big) \ov  \Gamma \big(\frac{1}{2}+ i \lm\big)}
\te + \psi^{(-2)} \big(\frac{1}{2}- i \lm\big) + \psi^{(-2)} \big(\frac{1}{2}+ i \lm\big)\ . 
\label{A.1.17}
\ee
Here $\psi^{(n)}(x)$  is the polygamma function, $\psi(x) = \psi^{(1)}(x)$,   and 
$A$ is the Glaisher constant,  
 \be\te   \ln A= \frac{1}{12}- \z'_R (-1)=\frac{1}{12}\big[ \g_E +\ln( 2 \pi) \big] -\frac{1 }{2\pi^2} \z'_R (2) 
= 1.282... \ . \la{gl}\ee
Thus $ \z_2^\ss (0) =- \frac{11}{2880}\,$   and 
\bea
{\te \z_2^{\ss\, \prime} (0) =- \frac{1}{144} \big[1 +   6 \ln (2 \pi) + 24 \ln A \big] + I_1}\,, 
\qquad\qquad 
I_1 \equiv  4  \int_0^{\infty} d \lm\  \frac{\lm}{e^{2 \pi \lm}+1} \ {\cal H} (\lm)\ . 
\label{A119}
\ee
It is not clear how    to compute  the convergent integral $I_1$ in ~\eqref{A119} analytically, 
but it 
 is   straightforward to evaluate it numerically:
 \be    I_1 = 0.117854... \ . \la{A120} \ee
Combining~\eqref{A.17} and~\eqref{A119} we finally get $ \z^\ss (0) =-\frac{1}{90}$  and 
\be
 \te
  \z^{\ss\, \prime} (0) = - {1\ov 2880}\big[41+ 114 \ln 2 +  120 \ln  \pi + 480 \ln A    + 5040\,  \z'_R (-3) \big] +  I_1\,,
\label{120}
\ee
where $ \z'_R (-3)= 0.005...$. 
%

In  the Majorana fermion  case \rf{a4}, \rf{a5}   we get 
\be 
&&K^\ff (t) =  -{\te \frac{1}{4 \pi^2}} \sum_{\ell=0}^{\infty} (2 \ell+2) e^{-t (\ell +1)^2}
\int_{0}^{\infty} d \lm\, \lm\, \coth(\pi \lm) 
e^{-t \lm^2 }\,,\\
&&  \z^\ff (z) =   2 \sum_{\ell=0}^{\infty} (2 \ell+2) 
\int_{0}^{\infty} d \lm\,  \frac{\lm \coth (\pi \lm)}{ [\lm^2 + (\ell+1)^2]^z}\,. 
\label{A.2.2}
\ee
The computation of $ \z^\ff (z) $ follows similar steps as  above, i.e. splitting 
$\coth (\pi \lm)= 1 +{2}({e^{2 \pi \lm}-1})^{-1}$, etc. We find that  $\z^\ff (0)= - \frac{11}{180}\,$   and\foot{Here we used that 

$
\sum_{r=2}^{\infty}  (-\lm^2)^r  r^{-1}  \z_R (2 r -1) =  \g_E \lm^2 + \ha 
\int_0^{\lm^2} d x\, \big[\,  \psi (i \sqrt{x} ) + \psi ( - i \sqrt{x})\big]
$.}
\bea
&&
\z^{\ff\,  \prime} (0) ={\te {1 \ov 180}} \big[7 - 120 \ln A - 720\, \z'_R (-3)\big]  +  2 I_2\,, \ \qquad  \
I_2 \equiv 4 \int_0^{\infty} d \lm \frac{\lm}{e^{2\pi \lm}-1} \, {\cal G} (\lm) \ , \ \ \ \  \ \la{nb}  \\
&&{\cal G} (\lm) = \ha \int_0^{\lm^2} d x\,  \Big[ \psi \big(i \sqrt{x} \big) + \psi \big( - i \sqrt{x}\big)\, \Big]
 =  i \lm\,   \ln{ \G (-i \lm) \ov  \G (i \lm)}  +  \psi^{(-2)} (-i \lm)+   \psi^{(-2)} (i \lm)\  .\ \   \ \ \ \ \ \la{bn}
\eea
The numerical  value of $I_2$ is 
\be   I_2 =    0.237101...   \ . \la{nbc}\ee
To   determine the vector contribution \rf{326} to $\zeta'(0)$ we need to  add  to the two scalar 
contributions  \rf{120} an extra term  corresponding to 
\be \la{kk} 
  \td K= -  \te \frac{1}{2\pi}K^\ss_{AdS_2}(t)   +  \frac{1}{2\pi} K^\ss_{S^2}(t)\equiv \tK_1 (t) + \tK_2 (t) \ , \ee
where $K^\ss_{S^2}$  and $ K^\ss_{AdS_2}$  were given in \rf{aa3} and \rf{3.9}. 
The  $\z$-function corresponding to the  $AdS_2$   part $\tK_1(t) $   is 
\be 
&& \td{\z}_1 (z)=2  \int d\lm\,  \frac{\lm\,  \tanh (\pi \lm)}{ \big(\lm^2+ \frac{1}{4}\big)^z}\,,\ \ \ \qquad  \ \ \ \ 
 \td{\z}_1 (0) =- \te{ 1\ov 3}\ ,\la{zx}  \\
&&
 \td{\z}'_1 (0) ={\te \frac{1}{6}  }(1 +\ln 2)  -2 \ln A + 2 \int_0^{1/4} dx\ \psi \big(\sqrt{x}+\ha \big)
={\te {1\ov 6} }-   \ln (2 \pi) + 4 \ln A \,. \quad
\label{A33}
\ee
The $S^2$  part   contains  the IR singular zero-mode ($\ell=0$)  contribution. It may be regularized  by adding 
a small mass  parameter $\mm^2$, 
\be \td{K}_2 (t)={\te  \frac{1}{8\pi^2} }\sum_{\ell=0}^{\infty} (2 \ell +1)\, e^{-s [\ell (\ell+1) + \mm^2]}\,,  \qquad \ \ 
 \td{\z}_2 (z) =
-\sum_{\ell =0}^{\infty} \frac{ 2 \ell+1}{ \big[ \big(\ell +\frac{1}{2}\big)^2 +(\mm^2 -{1\ov 4})\big]^z} 
\,. 
\label{A.3.11}
\ee
As a result, 
\be 
 \td{\z}'_2 (0)= {\te \frac{1}{6}} (1 +  \ln 2)   - 2  \ln A 
-   \int_0^{ -1/4+\mm^2} d x \big[ \psi \big(\ha  +i \sqrt{x} \big) + \psi \big(\ha  - i \sqrt{x}\big) \big] \,.
\label{A3.1}
\eea
Taking the limit $\mm\to 0$ and omitting the singular $\ln \mm$ term  we get  
\be 
 \td{\z}'_2 (0) =\te \frac{1}{6} + 4 \ln A + \ln \mm^2 \ \to \ \te \frac{1}{6} + 4 \ln A = \frac{1}{2}- 4 \z'_R (-1)\,. 
\label{A12}
\ee
This is the same  expression as found in 
\cite{Weisberger:1986qd}. 
Note that  including  formally (the  regularized)   zero-mode contribution 
 in the  $\z $-function 
gives  $\td{\z}_2 (0)= - {1\ov 3}$  which  matches the value of the corresponding  Seeley coefficient. 
Summing up  the  expressions for  $\td \z_1$ and  $\td \z_2$  in \rf{A33} and \rf{A12} we get   
$\td{\z} (0) =- \frac{2}{3}$   and 
\be \te 
 \td{\z}' (0) = \frac{1}{3} -\ln (2 \pi) + 8 \ln A \,. 
\label{A34}
\eea
We can  now  combine the above results  to  find  the total values of $\z(0)$ and $\z'(0)$  for the 
$\N=4$   supersymmetric  Abelian  gauge theory in \adt background.
Explicitly, for  the sum of the gauge field $\z^\gv = \z^\vv -  2 \z^\ss = 2 \z^\ss + \td \z$ (cf. \rf{326}), 
6 real scalar $\z^\ss$ and 4 Majorana fermion $\z^\ff$  contributions, i.e.  
$\z =\z^\gv  +  6 \z^\ss  +    4 \z^\ff =  8 \z^\ss + 4 \z^\ff   + \td \z$,   we get 
\be 
  \z (0)&=&  \te 8 \z^\ss (0) +    4 \z^\ff (0) + \td{\z} (0)=    - 8\times  \frac{1}{90} - 4 \times   \frac{11}{180}-  \frac{2}{3}= -1 \ , \la{zez}\\
\z'   (0) &=& 8 \z'^\ss (0) +4 \z'^\ff (0) + \td{\z}' (0)
\nonumber \\
  &=&  \te - {1 \ov 120} \big(   -45 + 158 \ln 2 + 160 \ln \pi\big)  +4  \ln A  - 30 \z'_R (-3)
  + 8 (I_1  +  I_2) \,. 
\label{A27}
\eea
The integrals  $I_1$ and $I_2$   were   given  in \rf{A119}, \rf{A120}  and   \rf{nb}, \rf{nbc}. 

We conclude   that the value of $\z(0)$ is the same as of $B_4$  in \rf{v} and  the  constant  parameter-independent 
 part $C_1$  of the 1-loop effective action   \rf{03} is   proportional to \rf{A27} (cf. \rf{z}). Its numerical value is  
\be C_1= - \ha \z'(0) =  -0.809684...  \ . \la{c1}\ee

\section{The \adss string  one-loop correction to circular  Wilson loop revisited}
\def\theequation{B.\arabic{equation}}
\setcounter{equation}{0}
\def \adn {{AdS_2}}

\def \tG  {{\bar  \G}}

For completeness,   let us review     the expression for the \adss string  one loop correction to  the  ($k$-wound) circular 
Wilson loop \ci{ber,dgo} found in  the
   limit when one first  takes $N = \infty$    and  then expands in large   string tension $\sql$  for fixed $k$.  
Here one represents  the planar 
expectation   value  $ \langle W \rangle = e^{-\tG}$, \ $\tG=\tG_0 + \tG_1 +  \tG_2 + ... $, 
 by the  string path integral with a disc-like world sheet ending on a circle at the boundary of \adss  space. 
  The classical  world-sheet  metric  is  that of the ($k$-wrapped version of)  
   euclidean $AdS_2$   with $S^1$  boundary  
    and so the  classical action (proportional to  the renormalized  volume in   \rf{voo})  is 
$\tG_0= - k\sql$,     while   $\tG_1=\G_1(k), \ \tG_2 = {1\ov \sql} \td \G_2(k),$ etc. 

The general form of the conformal-gauge string one loop   correction   was given   in \cite{dgt}  as 
\be 
\tG_{1}=  \ha  \ln \frac{[{\rm det} (-\nabla^2 +2  )]^3 \ [{\rm det} (-\nabla^2 )]^5}{[{\rm det} (-\nabla^2 +\frac{1}{4}R +1) ]^8}\,. 
\label{bb1}
\ee
In the straight  line WL case the classical action and loop corrections  are proportional to the volume 
of $\widetilde{AdS}_2$   with   boundary  $R$  which has zero renormalized value  \rf{voo}  and thus 
they    should  be  assumed to   vanish   \ci{dgt,kt},   in agreement with $\langle W \rangle =1$ on the gauge theory side.  

In the circular case the   computation   of the corresponding determinants   was    carried out  using different methods in \ci{dgt,kt,km}, 
with the finite  part    of the resulting expression for $\tG_1(k)$   being  \ci{kt}
\be 
\tG_{\rm 1\,fin} (k) =  \tG_{\rm1\, fin} (1) +  ( 2 k + \ha) \ln k  - \ln  k! \ , \ \ \ \ \ \ \ \ \ \quad
\tG_{\rm1\, fin} (1) = \ha \ln (2 \pi) \ .   \la{bb2} \ee
At the same  time, the gauge theory   expression \rf{b2}  taken at  $N\to \infty$ and then expanded in
 large $\l$ for fixed $k$ gives 
the familiar modified Bessel function expression \ci{Erickson:2000af,Drukker:2000rr,Drukker:2005kx}
$\langle W \rangle = 2  ( k \sql)^{-1}  {I}_1 ( k \sql)=  e^{k\sql - \G_1 + ...} $,  where 
\be \te \G_1 = \hat  \G_1   + { 3 \ov 2} \ln k + { 3 \ov 2} \ln \sql  \ , \ \ \ \ \ \ \ \ \ \ \ \ 
\hat \G_1 = \ha \ln { \pi \ov 2}\ .  \la{bb3} \ee
Here the second ${ 3 \ov 2} \ln k$ term is  of course the same  as the one present in \rf{33}. 
The third  term  may be attributed to   the presence of the  string tension 
 normalization factor for  the three  (Mobius-symmetry) 
  ghost zero modes on the disc \cite{Drukker:2000rr}, which was 
not included in \rf{bb1},\rf{bb2}. Even ignoring this term,  $\tG_{\rm1\, fin} (1)$ in \rf{bb2} still differs   from 
$\hat \G_1  $ in \rf{bb3}  by an extra $\ln 2$ term. 
This   difference    may be coming from  a numerical  factor  in   normalization of  the  disc zero modes 
or   from the ratio
of the ghost and  the two longitudinal mode determinants  (assumed to be  equal to  one 
 in \rf{bb1}) once they are  computed   with proper boundary 
conditions (cf. \ci{km}). 

A resolution of  this  problem    may   lead to a   change in 
 the $k$-dependence of  $\G_{\rm 1\,fin} $  in \rf{bb2} 
   making  it match the second term in the 
gauge-theory result  \rf{bb3}.  
Here we will  not attempt to resolve this issue 
  and   will only   review  and complete  the  original computation  in  \ci{dgt} 
 of $\G_{\rm1\, fin}  $  in \rf{bb2}  for $k=1$
  based   on 
expressing   the determinants in \rf{bb1}  in terms of the  
known  \ci{cam,camp} heat kernels  of  the scalar and spinor Laplacians on  $AdS_2$.
This  provides  an alternative  to the  derivation  of $\G_{\rm1\, fin}  $    in \ci{kt}.

The  discussion   below  repeats   the  computation  in Appendix B.1 in \ci{dgt}.    There 
it was  assumed to apply to the straight string case  and it was not appreciated
 that in the case of $\widetilde{AdS}_2$ 
with the boundary $R$ the result, proportional to the volume $V_{\widetilde{AdS}_2} \sim {1\ov \vee}$ contains only  IR divergent piece 
that may be discarded \ci{kt}.
  The expressions in Appendix B.1 in \ci{dgt}
are, in fact,  
  literally valid  in the circular string case 
where the  classical world sheet metric is that of $AdS_2$  with  $S^1$  boundary   which has finite renormalized
 volume \rf{voo}.\foot{The  expressions
 for heat kernels  and $\zeta$-functions  there did not  contain  the   volume factor 
 $V_{AdS_2}=-2 \pi$  which we  will include below.  We will also   perform the final  step of  summation of the bosonic and   fermionic contributions  that was not done explicitly in  Appendix B.1 in  \ci{dgt}.}
 
$\tG_1$   in \rf{bb1}    contains  the  contributions   of   3 scalars  with  mass-squared  $m^2=2$, 5 scalars with $m^2 =0$ and
 8 Majorana fermions with $m^2=1$, 
propagating  in $AdS_2$.  
We  will  set the $AdS_2$
radius to  1  since $ \tG_{1}$   does  not depend on it
(assuming   UV divergences  eventually  cancel \cite{dgt}).
  As in \rf{a1}   here 
\be \la{ab1} 
\ha \ln \det \Delta = - \ha   V_{{AdS_2}} \int_\ep^\infty { dt \ov t} \  K(t) \ , \ \ \ \ \ \ \ \ \ \ \ 
\  V_{{\adn}} = - 2\pi \  . 
\ee
The   trace of  heat  kernel   for massive   scalars and  fermions  may be written as 
\be 
&& K(t) = \frac{1}{2 \pi} \int_0^{\infty} d \lm\ \mu (\lm) \ e^{- t (\lm^2 +M)}\,, 
\label{4.7}\\
&&
\mu (\lm) =\lm \tanh (\pi \lm)\,, \qquad M= \te \frac{1}{4}+ m^2 \ \ {\rm for} \ {\rm scalar}\,, 
\la{4.88} \\
&&
\mu (\lm) =- \lm \coth (\pi \lm)\,, \qquad M= m^2 \ \ \ \ \  {\rm for} \ {\rm  Majorana\ fermion}\,. 
\label{4.8}
\eea
The corresponding $\z$-function is (cf. \rf{4.1})
\be 
\z (z) = - { 1 \ov \G(z) } \int_0^{\infty} d \lm\,  \mu (\lm) \int_0^{\infty} d t\ t^{z-1} \ e^{-t(\lm^2+M)}= -
 \int_0^{\infty} d \lm\ \frac{\mu (\lm)}{(\lm^2+M)^z}\,. 
\label{4.10}
\ee
%
%
Starting with the scalar case   we may  split the spectral density  as 
$\tanh (\pi \lm)= 1- {2}({e^{2 \pi \lm}+1})^{-1}\,.$  
The contribution from the first term to $\zeta' (0, M)$ is then 
$
-\ha  M (\ln M -1)\,. 
$
The second term  gives an exponentially convergent integral for large $\lm$  so we can set $z=0$ inside the integral.
Then the  scalar   $\z$-function is 
\be 
\z^{\ss\, \prime}  (0, M)=- {\te \frac{1}{2 }} M (\ln M -1) -  J_1(M) \ , \ \ \ \ \ \ \ \ \  
 J_1(M)\equiv   2  \int_0^{\infty} d \lm \ 
\frac{\lm\, \ln (\lm^2 +M)}{e^{2 \pi \lm}+1}\,. 
\label{4.14}
\ee
We  may write  $J_1 (M) =\int_0^{M} d x\  \frac{\pt J_1 (x)}{\pt x} + J_1 (0)$ where \ci{camp}
\be 
\frac{\pt J_1 (M)}{\pt M}
&=&
2  \int_0^{\infty} d\lm\ \frac{\lm}{(e^{2 \pi \lm}+1) (\lm^2+M)} =  -\ha  \ln M
+   \psi \big(\sqrt{M}+\ha \big)  \,, 
\label{4.17} \\ 
J_1 (0)&=& 2 \int_0^{\infty} d \lm\  \frac{\lm\, \ln \lm^2 }{e^{2 \pi \lm}+1}=  {\te {1 \ov 12}} ( 1+ \ln 2 ) -  \ln A
\,.
\label{4.19}
\ee
Thus 
%
\be 
\z^{\ss\, \prime} (0, M)=   - {\te {1 \ov 12}} ( 1+ \ln 2 ) +  \ln A    -  \int_0^{M} dx\ \psi \big(\sqrt{x}+\ha \big) \,, 
\label{4.21}
\ee
where $A$ is the Glaisher constant \rf{gl}.  
 The total bosonic contribution of 3  scalars     (see \rf{4.88})  with 
  $M={1\ov 4} + 2={9\ov 4} $  and 5  scalars with  $M={1\ov 4} $  is  found  (after doing the integrals)  to be 
%
\be 
\zeta'_B (0) =  3 \z^{\ss\, \prime} \big(0, \te\frac{9}{4}\big) + 5 \z^{\ss\, \prime} \big(0, \frac{1}{4}\big) 
  =- \frac{20}{3} + 7 \ln (2 \pi) - {16 \ln A}\,. 
\label{4.22}
\ee
Performing a  similar   computation  in the fermionic case \rf{4.8} (with 
$\coth (\pi \l)= 1+ {2}({e^{2 \pi \l}-1})^{-1}$, etc.) we get 
\be 
&&\z^{\ff\, \prime}  (0, M)=\ha  M (\ln M -1)   -  J_2  (M) \ , \ \ \quad \ \    J_2 (M)= 2  \int_0^{\infty} d \lm \
\frac{\lm\, \ln (\lm^2 +M)}{e^{2 \pi \lm}-1}\,, \ \ \ \ 
\label{4.25}\\  &&
\frac{\pt  J_2 (M)}{\pt M} = 2  \int_0^{\infty} d\lm\  \frac{\lm}{(e^{2 \pi \lm}-1) (\lm^2+M)}=
\ha  \ln M -   \ha  (\sqrt{M})^{-1}  -  \psi (\sqrt{M})\,.
\label{4.27}\\
&&  \z^{\ff\, \prime}  (0, M)=  - {\te { 1 \ov 6} }  + 2 \ln A      + {\sqrt{M}} + \int_0^M d x\ \psi(\sqrt{x})\,. 
\label{4.30}
\ee
The  contribution of  8 Majorana fermions with $M=m^2=1$ is then 
\be 
\z'_F (0) = 8  \z^{\ff\, \prime} (0,1) = \te \frac{20}{3 } -  {8 \ln (2 \pi)} + {16 \ln A}  \,. 
\label{4.31}
\ee
%
%
Thus the total 1-loop  correction  coming from \rf{bb1}  is 
\be 
\G_{1}=  - \ha   \z'_B (0) -\ha  \z'_F (0)  =\ha  \ln (2 \pi)\,, 
\label{4.32}
\ee
which is 
indeed the same as $\G_{\rm 1\, fin}$ in \rf{bb2}.

\

It  should be   possible  to generalize the   above computation to the case of $k$-wrapped    circular  Wilson loop
to check the result \rf{bb2}  of \ci{kt}.  The corresponding solution in $AdS_3$  with the metric $ds^2 = z^{-2} ( dr^2 + r^2 d \phi^2 + dz^2 )$ 
is  described  by $ \phi = k \tau, \  z= \tanh   k \s, \ r= \cosh^2 k \s $, $z= \sqrt{1- r^2}$,  where $\tau \in (0, 2 \pi), \ \s\in (0, \infty)$. The  induced  metric  is 
$ds^2 =   k^2 ( \sinh\k\s)^{-2} ( d\tau^2 + d \s^2) = d \xi^2 + k^2  \sinh^2 \xi\,  d \tau^2$, where 
$e^\xi = \tanh {k\s\ov 2}$. For $k=1$ 
 this is  the standard  regular  $AdS_2$ metric   but  for  general $k$ it has 
a conical singularity   at $\xi=0$   with negative deficit $\delta= 2 \pi (1-k), \ k=2, 3, 4, ...$.  
While we are interested in the case of  integer $k$ let us formally consider $k$  as an arbitrary real number. 
Then setting  it   to be 
 $1/n$  with an integer $n$    we get  an orbifold $AdS_2/Z_n$ (with a conical singularity of positive deficit)  and the corresponding 
heat kernel  can be found as a sum over images  \ci{ms,jon,gul}.\foot{Explicitly, it is given by  the  following modification of  \rf{xa}:\\
$K_{AdS_2} (x,x';t) ={\te  { 1 \ov 2 \pi}}\sum_{r=0}^{n-1} \int^\infty_0  dv \,  v \, \tanh ( \pi v)\, P_{- {1\ov 2} + i v } (\cosh \ddd(\xi,\tau; \xi', \tau'+ { 2 \pi r\ov n}))
\ e^{ - t ( v^2 + {1\ov 4})} 
\ $.\\
Similar construction applies in the spinor  case \ci{gul}.
}
In the case of   a  cone of a 2-plane  $ds^2 = d \xi^2 + k^2  \xi^2   d \tau^2$   with  generic  $k$   the heat 
  kernel can be found using ``re-periodisation"  trick 
\ci{dowk} and the same idea applies  to the cone of $AdS_2$  with explicit expression for the scalar  case given in  \ci{ms}. 
The   analogous  expression   for  the  spinor   heat kernel for generic $k$ 
 can be   found using  the results of  \ci{camp,camf}  and following the  examples  of the 
 cones of  2-plane and 2-sphere in  \ci{fm}. 
We leave  a  detailed computation for an  integer $k$  for the future.


\end{document}